# Stationary Bound States of Dirac Particles in the Schwarzschild Gravitational Field

M.A. Vronsky, M.V. Gorbatenko, N.S. Kolesnikov, V.P. Neznamov[1], E.Yu. Popov, I.I. Safronov

RFNC-VNIIEF, 37 Mira Ave., Sarov, 607188, Russia

## Abstract

Nondecaying bound states of elementary spin - half particles are validated and calculated numerically for the Schwarzschild gravitational field using a self-conjugate Hamiltonian with a flat scalar product for any value of the gravitational coupling constant.

Hilbert condition $g_{00} > 0$ leads to a boundary condition such that components of the vector of current density of Dirac particles are zero near the "event horizon".

At small values of the coupling constant, the energy spectrum is close to the hydrogen-like spectrum.

Based on the results of this study, we can assume that there exists a new type of collapsars, for which the Hawking radiation mechanism is not present. From the standpoint of cosmology, if the value of the gravitational coupling constant is small, $\alpha \ll 1$, the new type of nonradiating relict collapsars can manifest itself only through gravitation. Thus, they are good candidates for the role of "dark matter" carriers. In the wide range of admissible masses, there can exist collapsars of a new type with the masses of hypothesized WIMP particles, which are treated as representatives of "dark matter" in numerous scenarios of the expansion of the universe. The results of this study can lead to revisiting some concepts of the standard cosmological model related to the evolution of the universe and interaction of collapsars with surrounding matter.

[1] E-mail: neznamov@vniief.ru

## 1. Introduction

There are four solutions of the general theory of relativity that are known today for spherically and axially symmetric electrically charged and uncharged point-mass collapsars. They are the Schwarzschild metric [1], the Reissner-Nordström metric [2], the Kerr metric [3], and the Kerr-Newman metric [4].

The classical Schwarzschild solution is characterized by a spherically symmetrical point source of gravitational field of mass $M$ and radius of "event horizon" (gravitational radius)

$$r_0 = \frac{2GM}{c^2}. \tag{1}$$

In (1), $G$ is the gravitational constant, and $c$ is the speed of light. In the classical case, a particle measured by a clock of a distant observer reaches the "event horizon" in an infinitely long time.

For a probe particle of mass $m$, the dimensionless gravitational coupling constant equals

$$\alpha = \frac{GMm}{\hbar c} = \frac{Mm}{m_p^2} = \frac{r_0}{2l_c}. \tag{2}$$

In (2), $\hbar$ is the Planck constant, $m_p = \sqrt{\frac{\hbar c}{G}} = 2,2 \cdot 10^{-5}$ g is the Planck mass, and $l_c = \frac{\hbar}{mc}$ is the Compton wavelength.

As distinct from the interaction constants in the Standard Model, in the gravitational case, the coupling constant $\alpha$ can reach very high values. For an electron, the value $\alpha \approx 1$ characterizes a gravity source of mass $M = 0,5 \cdot 10^{15}$ kg. Then, for example, the gravitational interaction between an electron and a source of mass $M = M_{\odot} \simeq 2 \cdot 10^{30}$ kg will be characterized by $\alpha \simeq 4 \cdot 10^{15}$.

At present we know that in the centers of galaxies there are collapsars, the mass of which reaches billions of solar masses. In this case, the gravitational coupling constant of electron interaction with such objects will be $\alpha \simeq 10^{25}$.

In the Reissner-Nordström solution, a Schwarzschild spherically symmetric point source possesses an electric charge $Q$. The Kerr solution corresponds to a rotating Schwarzschild source with an angular momentum $\mathbf{J} = Mc\mathbf{a}$. In the Kerr-Newman solution, a rotating Kerr source possesses an electric charge $Q$.

In addition to the above solutions of the general theory of relativity, there exist metrics derived by coordinate transformations of the basic metrics.

For the Schwarzschild field, the following solutions can be noted: the Eddington-Finkelstein metric [5], [6], the Painleve-Gullstrand metric [7], the Lemaitre-Finkelstein metric [6], and the Kruskal metric [8].

Note the Doran solution [9] for the Kerr field.

Despite the evident electromagnetic analogy in atomic physics, bound states of Dirac particles in collapsar fields have been investigated comparatively scantily. For the gravitational case, bound states are believed to have complex energies. In this case, these states decay exponentially with time. The existence of resonant Schwarzschild states for massive scalar particles is discussed in [10] - [13] using the Klein-Gordon equation. The same problem for massive Dirac particles is considered in [14] - [18]. In these studies, a hydrogen-like spectrum with relativistic corrections is obtained for $\alpha \ll 1$ by direct solution of the Dirac equation in a weak Schwarzschild field for the real part of energy. In [19], the authors consider the problem of bound states in the Schwarzschild field using the Painleve-Gullstrand metric. Energy spectra with complex energies are obtained for $\alpha \ll 1$ and $\alpha \sim 1$. A hydrogen-like spectrum is also obtained for the real part of the energy at $\alpha \ll 1$ with different relativistic corrections as compared with [16].

In [20] - [22], authors developed a method for deriving self-conjugate Dirac Hamiltonians with a flat scalar product of wave functions within the framework of pseudo-Hermitian quantum mechanics for arbitrary, including time dependent, gravitational fields.

It follows from single-particle quantum mechanics that if the stationary Hamiltonian is Hermitian, if there are quadratically integrable wave functions, and if corresponding boundary conditions are specified, such Hamiltonians should provide for the existence of stationary bound states of particles with a real energy spectrum. In the current work, we explore such states of Dirac particles in the Schwarzschild field.

## 2. Analysis of the possibility of existence of bound states of spin –half particles in Schwarzschield gravitational field

We will use the system of units $\hbar = c = 1$, signature

$$\eta_{\underline{\alpha\beta}} = diag\left[1, -1, -1, -1\right] \tag{3}$$

and notation $\gamma^{\alpha}$, $\gamma^{\underline{\alpha}}$ for global and local Dirac matrices, respectively. As local matrices we use matrices in the Dirac-Pauli representation.

We first present a stationary self-conjugate Dirac Hamiltonian for the Schwarzschild metric as defined in [22].

The Schwarzschild metric is

$$ds^2 = f_s dt^2 - \frac{dr^2}{f_s} - r^2\left(d\theta^2 + \sin^2\theta d\varphi^2\right),$$
$$f_s = 1 - \frac{r_0}{r}. \tag{4}$$

In (4) we mean real values of $f_s > 0$ (the Hilbert condition: $g_{00} > 0$).

The self-conjugate Hamiltonian $H_\eta$ is

$$H_\eta = \sqrt{f_s} m\gamma^{\underline{0}} - i\sqrt{f_s}\gamma^{\underline{0}}\left\{\gamma^{\underline{1}}\sqrt{f_s}\left(\frac{\partial}{\partial r} + \frac{1}{r}\right) + \right.$$
$$\left. + \gamma^{\underline{2}}\frac{1}{r}\left(\frac{\partial}{\partial\theta} + \frac{1}{2}\mathrm{ctg}\theta\right) + \gamma^{\underline{3}}\frac{1}{r\sin\theta}\frac{\partial}{\partial\varphi}\right\} - \frac{i}{2}\gamma^{\underline{0}}\gamma^{\underline{1}}\frac{\partial f_s}{\partial r}. \tag{5}$$

## 2.1. SEPARATION OF VARIABLES

The Dirac equation with stationary Hamiltonian (5) allows the separation of variables, if the bispinor $\psi_\eta(\mathbf{r},t) = \psi_\eta(r,\theta,\varphi)e^{-iEt}$ is defined as

$$\psi_\eta(r,\theta,\varphi) = \begin{pmatrix} F(r) & \xi(\theta) \\ -iG(r) & \sigma^3\xi(\theta) \end{pmatrix} e^{im_\varphi\varphi} \tag{6}$$

and the following equation is used (see, e.g., [23])

$$\left[-\sigma^2\left(\frac{\partial}{\partial\theta} + \frac{1}{2}\mathrm{ctg}\theta\right) + i\sigma^1 m_\varphi \frac{1}{\sin\theta}\right]\xi(\theta) = i\kappa\xi(\theta). \tag{7}$$

In order to receive equality (7) we made an equivalent replacement of matrices in the Hamiltonian (5)

$$\gamma^{\underline{1}} \to \gamma^{\underline{3}}, \gamma^{\underline{3}} \to \gamma^{\underline{2}}, \gamma^{\underline{2}} \to \gamma^{\underline{1}} \tag{8}$$

In equalities (6), (7): $\xi(\theta)$ are spherical harmonics for the spin –half particles, $\sigma^i$ are two-dimensional Pauli matrices, $m_\varphi$ is the magnetic quantum number, and $\kappa$ is the quantum number of the Dirac equation:

$$\kappa = \pm 1, \pm 2 ... = \begin{cases} -(l+1), j = l + 1/2 \\ l, \quad j = l - 1/2 \end{cases}. \tag{9}$$

In (9), $j,l$ are the quantum numbers of the total and orbital momentum of a Dirac particle, respectively.

$\xi(\theta)$ can be represented as in [24].

$$\xi(\theta)=\begin{pmatrix} {}_{-1/2}Y_{jm_\varphi}(\theta) \\ {}_{1/2}Y_{jm_\varphi}(\theta) \end{pmatrix}=(-1)^{m_\varphi+1/2}\sqrt{\frac{1}{4\pi}\frac{(j-m_\varphi)!}{(j+m_\varphi)!}}\begin{pmatrix} \cos\theta/2 & \sin\theta/2 \\ -\sin\theta/2 & \cos\theta/2 \end{pmatrix}\times$$

$$\times\begin{pmatrix} \left(\kappa-m_\varphi+\frac{1}{2}\right) P_l^{m_\varphi-1/2}(\theta) \\ P_l^{m_\varphi+1/2}(\theta) \end{pmatrix}. \tag{10}$$

In (10), $P_l^{m_\varphi \pm 1/2}(\theta)$ are Legendre polynomials.

The separation of variables gives a system of equations for real radial functions $F(r),G(r)$. In the following these equations are written in dimensionless variables, $\varepsilon=\frac{E}{m},\ \rho=\frac{r}{l_c},\ \frac{r_0}{l_c}=2\alpha$.

## 2.2. EQUATIONS AND ASYMPTOTICS FOR RADIAL WAVE FUNCTIONS

A system of equations for real radial functions $F(\rho),G(\rho)$ is written as

$$\begin{aligned} &\left(1-\frac{2\alpha}{\rho}\right)\frac{dF}{d\rho}+\left(\frac{1+\kappa\sqrt{1-\frac{2\alpha}{\rho}}}{\rho}-\frac{\alpha}{\rho^2}\right)F-\left(\varepsilon+\sqrt{1-\frac{2\alpha}{\rho}}\right)G=0, \\ &\left(1-\frac{2\alpha}{\rho}\right)\frac{dG}{d\rho}+\left(\frac{1-\kappa\sqrt{1-\frac{2\alpha}{\rho}}}{\rho}-\frac{\alpha}{\rho^2}\right)G+\left(\varepsilon-\sqrt{1-\frac{2\alpha}{\rho}}\right)F=0. \end{aligned} \tag{11}$$

At $f_s>0$, the variable $\rho$ for the functions $F(\rho),G(\rho)$ is defined on the interval $(2\alpha,\infty)$. At $f_s>0$, Eqs. (11) show that, as in the classical case, quantum mechanics prohibits the presence of Dirac particles on and under the “event horizon”, $r\le r_0$ , i.e. $\rho\le 2\alpha$ .

Let us consider the asymptotics of solutions of Eqs. (11).

As $\rho\to\infty$ the leading terms of the asymptotics equal

$$\begin{aligned} &F=C_1e^{-\rho\sqrt{1-\varepsilon^2}}+C_2e^{\rho\sqrt{1-\varepsilon^2}} \\ &G=\sqrt{\frac{1-\varepsilon}{1+\varepsilon}}\left(-C_1e^{-\rho\sqrt{1-\varepsilon^2}}+C_2e^{\rho\sqrt{1-\varepsilon^2}}\right). \end{aligned} \tag{12}$$

In order to provide the finite motion of Dirac particles, we must use only exponentially decreasing solutions (12), i.e. in this case $C_2=0$ .

As $\rho\to 2\alpha$ $(r\to r_0)$,

$$F = \frac{A}{\sqrt{\rho - 2\alpha}} \sin\left(2\alpha\varepsilon \ln\left(\rho - 2\alpha\right) + \varphi\right),$$
$$G = \frac{A}{\sqrt{\rho - 2\alpha}} \cos\left(2\alpha\varepsilon \ln\left(\rho - 2\alpha\right) + \varphi\right). \tag{13}$$

In (12), (13), $C_1, A, \varphi$ are constants of integration.

The oscillating functions $F$ and $G$ in (13) are ill-defined at the "event horizon" but they are quadratically integrable at $\rho \neq 2\alpha$.

Hamiltonian (5) is Hermitian for the entire domain of $\rho$. We can show this using the general condition of Hermiticity of Dirac Hamiltonians in external gravitational fields proven in [20].

$$\oint ds_k \left(\sqrt{-g}\, j^k\right) + \int d^3x \sqrt{-g} \left[ \psi^+ \gamma^{\underline{0}} \left( \gamma^0_{,0} + \begin{pmatrix} 0 \\ 00 \end{pmatrix} \gamma^0 \right) \psi + \begin{pmatrix} k \\ k0 \end{pmatrix} j^0 \right] = 0. \tag{14}$$

For a stationary centrally symmetric Schwarzschild field $\gamma^0_{,0} \equiv \dfrac{\partial \gamma^0}{\partial x^0} = 0$, Christoffel symbols $\begin{pmatrix} 0 \\ 00 \end{pmatrix}, \begin{pmatrix} k \\ k0 \end{pmatrix} = 0$ and condition (14) reduces to

$$4\pi\rho^2 j^r \left(\rho \to \infty\right) + 4\pi\rho^2 j^r \left(\rho \to 2\alpha\right) = 0. \tag{15}$$

By definition and considering (8) the radial current density of Dirac particles is equal to

$$j^r = \psi^+_\eta f_s \gamma^{\underline{0}} \gamma^{\underline{3}} \psi_\eta. \tag{16}$$

Further we use functions (6) written as

$$\psi_\eta\left(\rho, \theta, \varphi\right) = \frac{1}{\rho} \frac{1}{\sqrt{f_s}} \begin{pmatrix} f(\rho) & \xi(\theta) \\ -ig(\rho) & \sigma^3 \xi(\theta) \end{pmatrix} e^{im_\varphi \varphi}. \tag{17}$$

As a result, the radial current density equals

$$j^r = \frac{i}{\rho^2} f(\rho) g(\rho) \left[\xi^+(\theta)\left(\sigma^3\sigma^3 - \sigma^3\sigma^3\right)\xi(\theta)\right] = 0,. \tag{18}$$

Thus, if we introduce a physically reasonable boundary condition at $\rho \to 2\alpha$, the system of equations (11) will possess a stationary real energy spectrum of bound states of spin-half particles.

## 3. DETERMINATION OF THE STATIONARY BOUND STATES OF DIRAC PARTICLES IN THE SCHWARZSCHILD GRAVITATIONAL FIELD

### 3.1. WEAK GRAVITATIONAL FIELD

In this case, the gravitational radius $r_0$ is much smaller than the Compton wavelength $l_c$ $(r_0 \ll l_c)$. Given that the distances $r \ll l_c$ cannot have any significant effect in quantum mechanics on the energy spectrum of Eqs. (11), we consider that the quantities $\frac{2\alpha}{\rho}$ can be ignored compared to unity over the entire range of variation of $\rho$. Then, Eqs. (11) take the form

$$\begin{aligned} &\frac{\partial F}{\partial \rho} + \frac{1+\kappa}{\rho} F - (\varepsilon + 1) G = 0, \\ &\frac{\partial G}{\partial \rho} + \frac{1-\kappa}{\rho} G + \left( \varepsilon - 1 + \frac{\alpha}{\rho} \right) F = 0. \end{aligned} \tag{19}$$

In weak fields, the value of $\varepsilon$ is close to unity and the summand $\frac{\alpha}{\rho} F$ in the second equation of system (19) should not be ignored. The solution of system (19) is close to the solution of a system of Dirac equations for the Coulomb potential

$$U(\rho) = \frac{\alpha_{em}}{\rho}, \tag{20}$$

$$\begin{aligned} &\frac{\partial F}{\partial \rho} + \frac{1+\kappa}{\rho} F - \left( \varepsilon + 1 + \frac{\alpha_{em}}{\rho} \right) G = 0, \\ &\frac{\partial G}{\partial \rho} + \frac{1-\kappa}{\rho} G + \left( \varepsilon - 1 + \frac{\alpha_{em}}{\rho} \right) F = 0. \end{aligned} \tag{21}$$

In (20), (21), $\alpha_{em}$ is the electromagnetic fine-structure constant.

Hence, the energy spectrum defined by Eqs. (19) is close to that of the Dirac equation for hydrogen-like atoms, if $\alpha_{em}$ is replaced with $\alpha$, and it can be written as

$$E_n \approx m \left( 1 - \frac{\alpha^2}{2n^2} \right). \tag{22}$$

If there are no relativistic corrections, this result coincides with the results obtained in [14] - [18].

The hydrogen-like spectrum (22) can also be obtained by turning in the nonrelativistic approximation from the Dirac equation with Hamiltonian (5) to a corresponding Schrödinger equation.

## 3.2. BOUNDARY CONDITION NEAR THE "EVENT HORIZON"

In order to numerically determine the spectrum of bound states in (11), we must specify a boundary condition for $\rho \to 2\alpha \ \left(r \to r_0\right)$.

Let us turn back to the form and values of current density components of Dirac particles for $\rho \to 2\alpha \ \left(r \to r_0\right)$.

Considering (10), (17),

$$j^{\theta} = \psi_{\eta}^{+}\sqrt{f_s}\gamma^0\gamma^1\psi_{\eta} = -\frac{2}{\rho^2\sqrt{f_s}} f\left(\rho\right)g\left(\rho\right)\left[\xi^{+}\left(\theta\right)\sigma^2\xi\left(\theta\right)\right] = 0, \qquad (23)$$

$$j^{\varphi} = \psi_{\eta}^{+}\sqrt{f_s}\gamma^0\gamma^2\psi_{\eta} = \frac{2}{\rho^2\sqrt{f_s}} f\left(\rho\right)g\left(\rho\right)\left[\xi^{+}\left(\theta\right)\sigma^1\xi\left(\theta\right)\right] \neq 0, \qquad (24)$$

The components $j^r$ (18) and $j^{\theta}$ (23) are equal to zero, the $\varphi$-component of the current (24) grows indefinitely as $\rho \to 2\alpha$.

Considering a domain of Dirac wave functions $\rho \in \left(2\alpha, \infty\right)$ the condition that there should be no particles on and under "event horizon" can be implemented if we suppose that components of the current density for $\rho \to 2\alpha \ \left(r \to r_0\right)$ are equal to zero.

Considering the form of functions (6), (10), (13), (17) and Eqs. (18), (23), (24), this condition equals

$$f\left(\rho\right)g\left(\rho\right)\Big|_{\rho \to 2\alpha} \to 0. \qquad (25)$$

From two possible versions of fulfilment (25) we will use the condition

$$g\left(\rho\right)\Big|_{\rho \to 2\alpha} \to 0. \qquad (26)$$

Some reason for this is a smallness of function $g\left(\rho\right)$ in comparison with $f\left(\rho\right)$ in nonrelativistic approximation of the Dirac equation.

Considering (17), (13) the condition (26) reduces to

$$\cos\left(2\alpha\varepsilon \ln\left(\rho - 2\alpha\right) + \varphi\right) \to 0 \text{ при } \rho \to 2\alpha \qquad (27)$$

$$\left(2\alpha\varepsilon \ln\left(\rho - 2\alpha\right) + \varphi\right)\Big|_{\rho \to 2\alpha} = \frac{\pi}{2}N, \quad N = \pm 1, 3, 5... \qquad (28)$$

Condition (28) determines the real energy spectrum of Eqs. (11), when these are solved numerically.

## 3.3. METHODS AND RESULTS OF NUMERICAL SOLUTION OF A SYSTEM OF DIRAC EQUATIONS FOR RADIAL WAVEFUNCTIONS IN THE SCHWARZSCHILD FIELD

### 3.3.1 NUMERICAL METHODS FOR DETERMINING EIGENVALUES AND EIGENFUNCTIONS OF THE DIRAC EQUATION IN THE SCHWARZSCHILD FIELD

For the functions $f(\rho)=\sqrt{\rho(\rho-2\alpha)}F(\rho)$, $g(\rho)=\sqrt{\rho(\rho-2\alpha)}G(\rho)$ (see (6), (17)), the system of equations (11) is written as

$$\begin{cases} (1-2\alpha/\rho)\dfrac{df}{d\rho}+\dfrac{\kappa\sqrt{1-2\alpha/\rho}}{\rho}f-\left(\varepsilon+\sqrt{1-2\alpha/\rho}\right)g=0 \\ (1-2\alpha/\rho)\dfrac{dg}{d\rho}-\dfrac{\kappa\sqrt{1-2\alpha/\rho}}{\rho}g+\left(\varepsilon-\sqrt{1-2\alpha/\rho}\right)f=0 \end{cases} \tag{29}$$

with boundary conditions

$$f(\rho_{\max})/g(\rho_{\max})=-\sqrt{\frac{1+\varepsilon}{1-\varepsilon}}; \quad g(\rho)\big|_{\rho\to 2\alpha}\to 0. \tag{30}$$

This problem can be solved in several ways. We consider two of them. The first is "inverse" numerical solution of system (29) starting from some $\rho_{\max}$ such that asymptotic behavior of (12), (30) is achieved (a natural choice for $\rho_{\max}$ is such that $\rho_{\max}\sqrt{1-\varepsilon^2}\gg 1$). Then we select $\varepsilon$ such that the solutions satisfy the boundary condition near the horizon. The second way is to introduce a new smooth function – a phase $\Phi$ coinciding with $\operatorname{arctg}(f/g)$ to within $k\pi$ $(k=1,2,3...)$.One can write a closed differential equation for this function such that asymptotic condition (12) and the condition at $\rho\to 2\alpha$ can be easily rewritten in phase terms.

If the Eqs. (29) is written as

$$\begin{pmatrix} \dfrac{df}{d\rho} \\ \dfrac{dg}{d\rho} \end{pmatrix}=\begin{pmatrix} a & b \\ c & d \end{pmatrix}\begin{pmatrix} f \\ g \end{pmatrix}, \tag{31}$$

then the phase equation has the following form:

$$\frac{d\Phi}{d\rho}=\frac{b-c}{2}+\frac{b+c}{2}\cos(2\Phi)+\frac{a-d}{2}\sin(2\Phi). \tag{32}$$

We introduce the variable $s=\dfrac{\rho}{2\alpha}-1$; the range of variation of $s$: $(0,\infty)$. Then the Eq. (32) with $a,b,c,d$ from the initial system of equations (29) has the following form:

$$\frac{d\Phi}{ds} = 2\alpha\left(\varepsilon\frac{s+1}{s} + \sqrt{\frac{s+1}{s}}\cos 2\Phi\right) - \kappa\frac{1}{\sqrt{s(s+1)}}\sin 2\Phi. \qquad (33)$$

In numerical calculations with $s \to 0$, it is convenient to introduce a variable

$$t = 2\alpha\varepsilon \ln(\rho - 2\alpha). \qquad (34)$$

Then, $s(t) = \frac{1}{2\alpha}e^{\frac{t}{2\alpha\varepsilon}}$ and Eq. (32) for the variable $t$ takes the form

$$\frac{d\Phi(s(t))}{dt} = 1 + s(t) + \sqrt{s(t)(1+s(t))}\frac{\cos 2\Phi}{\varepsilon} - \frac{\kappa}{2\alpha\varepsilon}\sqrt{\frac{s(t)}{1+s(t)}}\sin 2\Phi. \qquad (35)$$

In (35) the range of variation of $t$: $(-\infty, \infty)$.

Asymptotic condition (12), (30) in terms of phase equals

$$\Phi = -\text{arctg}\sqrt{\frac{1+\varepsilon}{1-\varepsilon}}, \text{ as } s \to \infty \qquad (36)$$

As a matter of fact, boundary condition (30), (27) as $\rho \to 2\alpha$ or $t \to -\infty$ is written as a condition for the phase:

$$\cos\Phi = \cos(t + \varphi) \to 0, \text{ as } t \to -\infty. \qquad (37)$$

Linear system (29) is solved implicitly by the Euler method with automatic step selection. Phase equations (32), (35) are solved numerically by the fifth-order Runge-Kutta implicit method with step control [25].

Let us briefly discuss the technology for determining the energy spectrum of the Dirac equations in numerical calculations.

The numerical solution of system (29) and phase equations (32), (35) are solved inversely from $\rho_{max}$ to $\rho \to 2\alpha$ or for the variables $s, t$ from $s_{max}$ to $t \to -\infty$. $\rho_{max}$ satisfies the condition $\rho_{max}\sqrt{1-\varepsilon^2} \gg 1$; $t_{min}$ in calculations is determined by the specified accuracy.

It follows from (35) that $\Phi = \varphi + t$ as $t \to -\infty$, where $\varphi$ is constant at given $\alpha, \kappa, \varepsilon$. Apparently the quantity $\varphi(\alpha, \kappa, \varepsilon)$ can be determined by solution of (35) with any specified accuracy. As a measure of accuracy we can use the degree of convergence of the phase in calculations to the straight line $\varphi + t$.

In the finite expression for the phase $\Phi$ at $t \to -\infty$ a constant summand $\varphi(\alpha, \kappa, \varepsilon)$ has total necessary physical information of a system under consideration: Schwarzschild collapsar – Dirac particle.

In the boundary condition (37) there is a variable $t \to -\infty$ except the well-determined quantity $\varphi(\alpha, \kappa, \varepsilon)$. It is necessary a fixation of $t_{min}$ for uniqueness of the boundary condition. It means that the boundary condition is formulated at some distance above the "event horizon". In this paper in Eq. (37) $t_{\min}$ equals

$$t_{\min} = -\pi K, \tag{38}$$

where $K$ is an arbitrary large integer. Then the condition (37) takes the form

$$\cos\varphi = 0 \tag{39}$$

$$\varphi(\alpha,\kappa,\varepsilon) = \frac{\pi}{2}N, \quad N = \pm 1, \pm 3, \pm 5... \tag{40}$$

Note that the phase $\varphi(\alpha,\kappa,\varepsilon)$ in (40) is determined by solution of Eqs. (33), (35) with any specified accuracy at $t \to -\infty$.

### 3.3.2 NUMERICAL SOLUTIONS OF THE DIRAC EQUATION IN THE SCHWARZSCHILD FIELD

#### 3.3.2.1 NUMERICAL DETERMINATION OF THE ENERGY SPECTRUM

When solving system of equations (29) or equivalent phase equations (32), (35) for the given values of $\alpha$ and quantum number $\kappa$ (9), we can obtain numerical functions $\varphi(\varepsilon)$. We denote the lowest energy level at fixed values of $\alpha$ and $\kappa$ as $\varepsilon_0$. If the values of $\alpha$ are small, the functions $\varphi(\varepsilon)$ at $\varepsilon < \varepsilon_0$ grow in a stepwise manner, while at $\varepsilon > \varepsilon_0$ the phase varies slightly, $\left(\Delta\varphi < \frac{\pi}{2}\right)$. If $\alpha$ increases, the functions $\varphi(\varepsilon)$ gradually become smoother.

If $\alpha$ increases and reaches $\alpha = \alpha_{cr}$, the maximum of the function $\varphi(\varepsilon)$ becomes equal to $N\frac{\pi}{2}$. This means that the lowest energy level $\varepsilon = \varepsilon_0^{cr}$ corresponds to this maximum. With further increase in $\alpha$, the maximum moves towards higher values of $1-\varepsilon$, growing in its magnitude.

In fact, for all the values of $\alpha$ at $\varepsilon \sim 0$, the phase value is close to zero.

Figs. 1 ÷ 4 show some functions $\varphi(\varepsilon)$ calculated for $\kappa = -1$.

Applying boundary condition (39) to the function $\varphi(\varepsilon)$, we obtain the sought discrete energy spectrum.

Several first lowest energy levels in Figs. 1÷4 are marked with an intersection of the phase curves $\varphi(\varepsilon)$ with horizontal lines $N\frac{\pi}{2}$ $(N = \pm 1, 3, 5...)$.

The above behavior of the function $\varphi(\varepsilon)$ leads to features in the functions $\varepsilon_n(\alpha,\kappa)$.

One can see in Figs. 1÷4 that the energy of the bound Dirac particle decreases noticeably as $\alpha$ increases. After $\varepsilon = \varepsilon^{cr}$ is reached, the particle energy drops and then decreases smoothly as $\alpha$ increases.

Near the maximum of the function $\varphi(\varepsilon)$ at $\alpha \geq 1$ there is a noticeable "energy desert" between energy levels (see Fig. 2,3).

Fig. 4 suggests that at $\alpha \geq 10$ a noticeable number of levels occur with a binding energy close to the total particle energy $\varepsilon = 1(E = m)$.

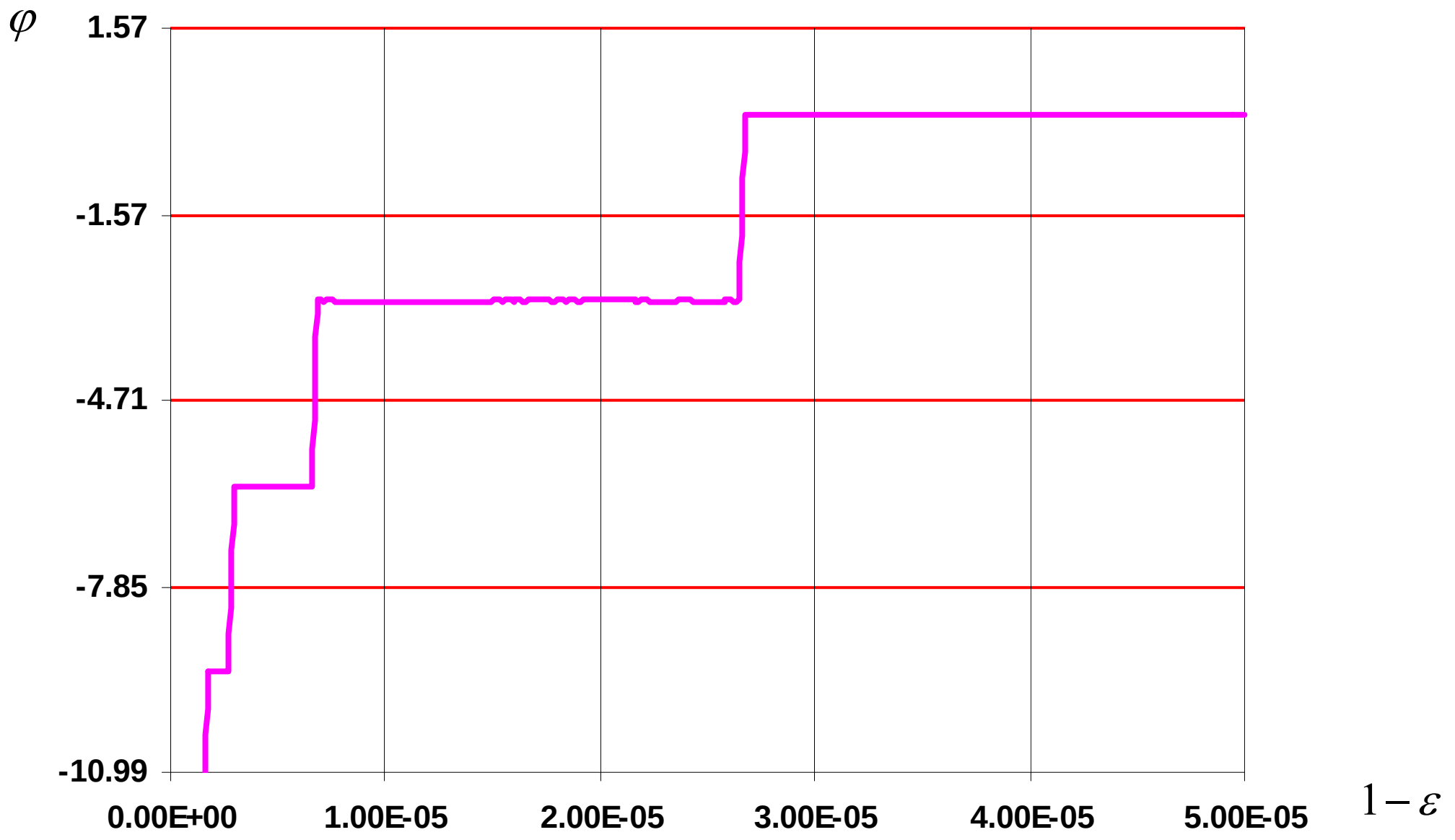

Figure 1. Phase behavior at $\alpha = 1/137$.

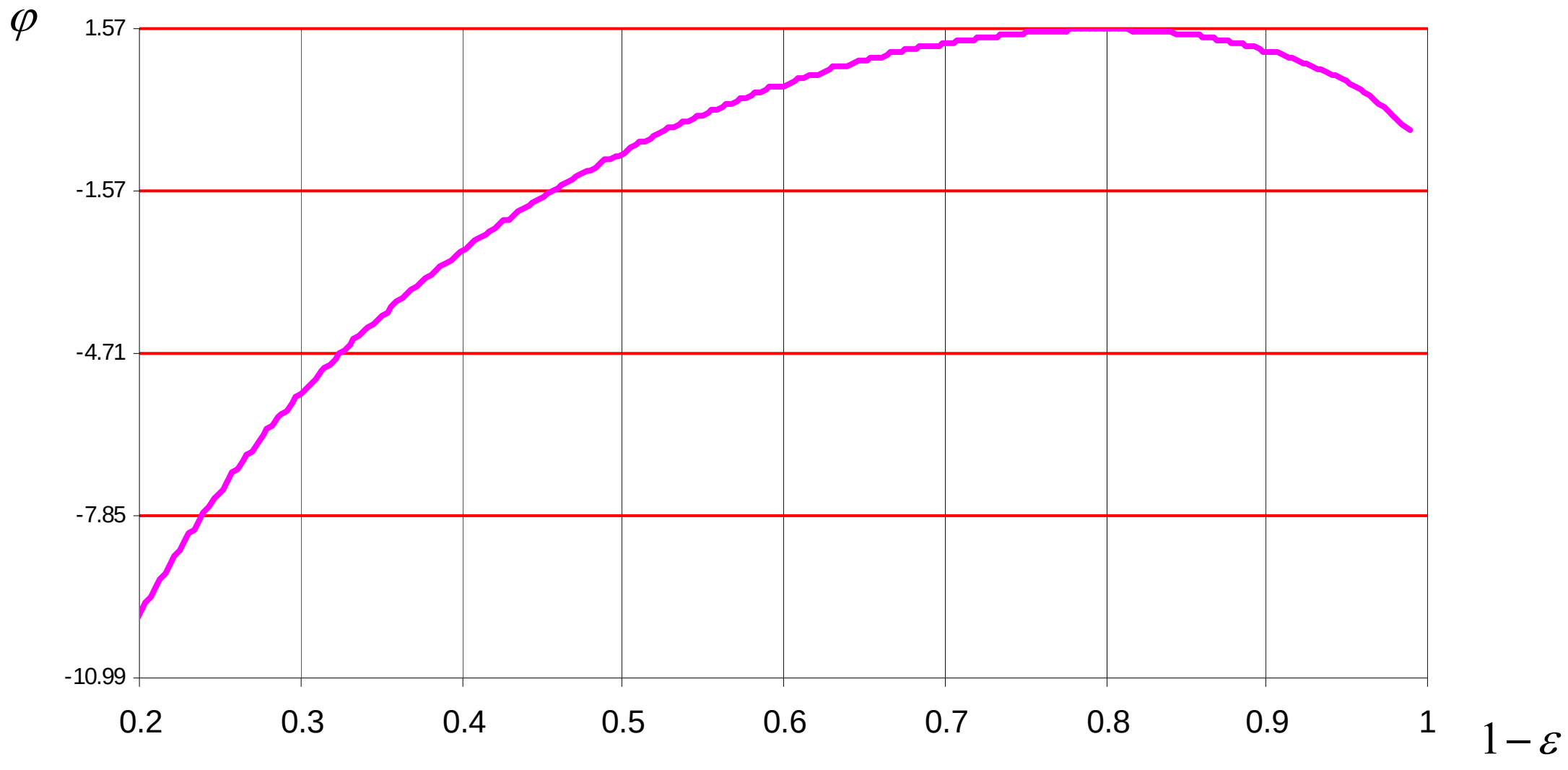

Figure 2. Phase behavior at $\alpha = 2{,}65$.

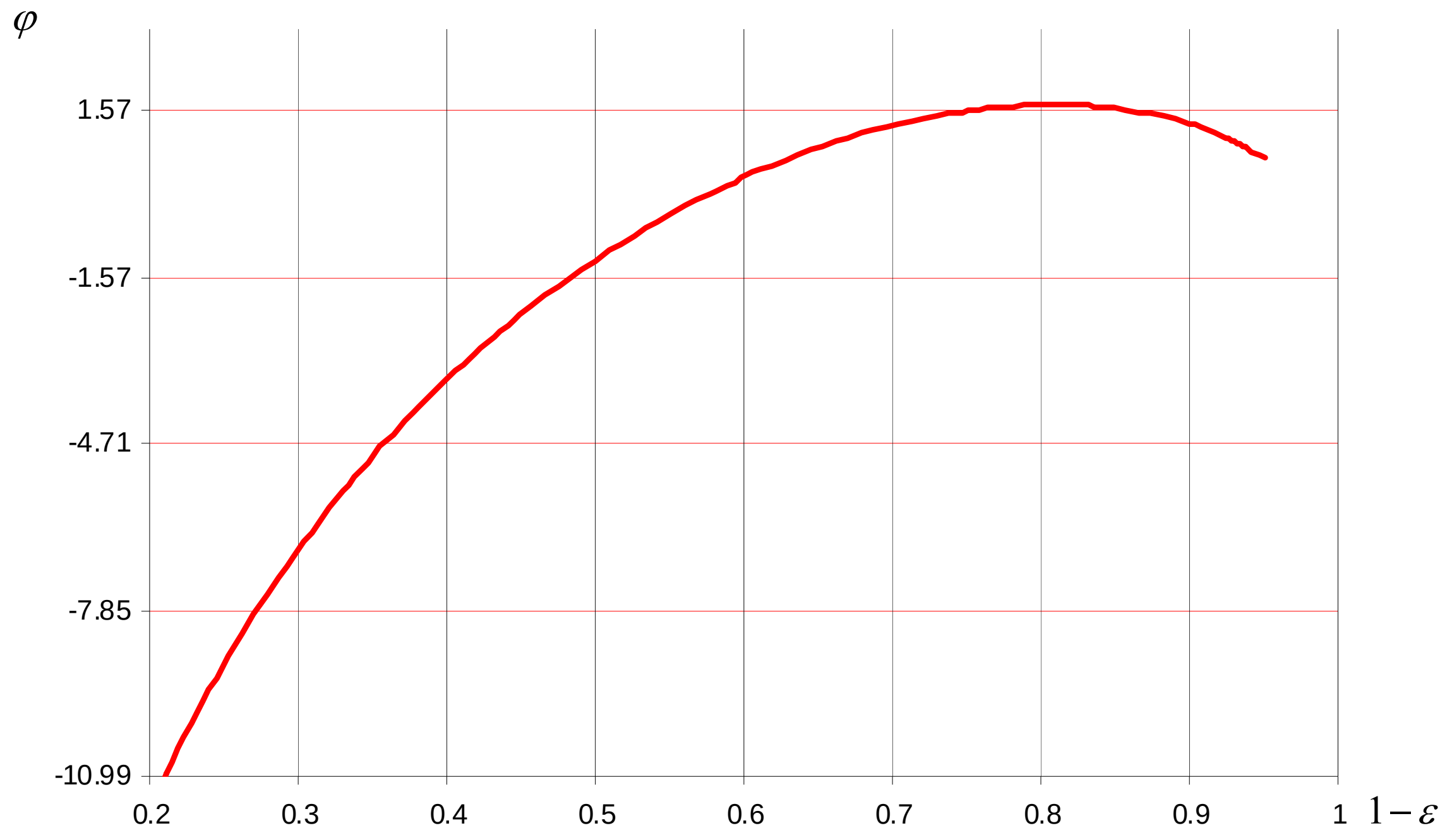

Figure 3. Phase behavior at $\alpha = 3$ .

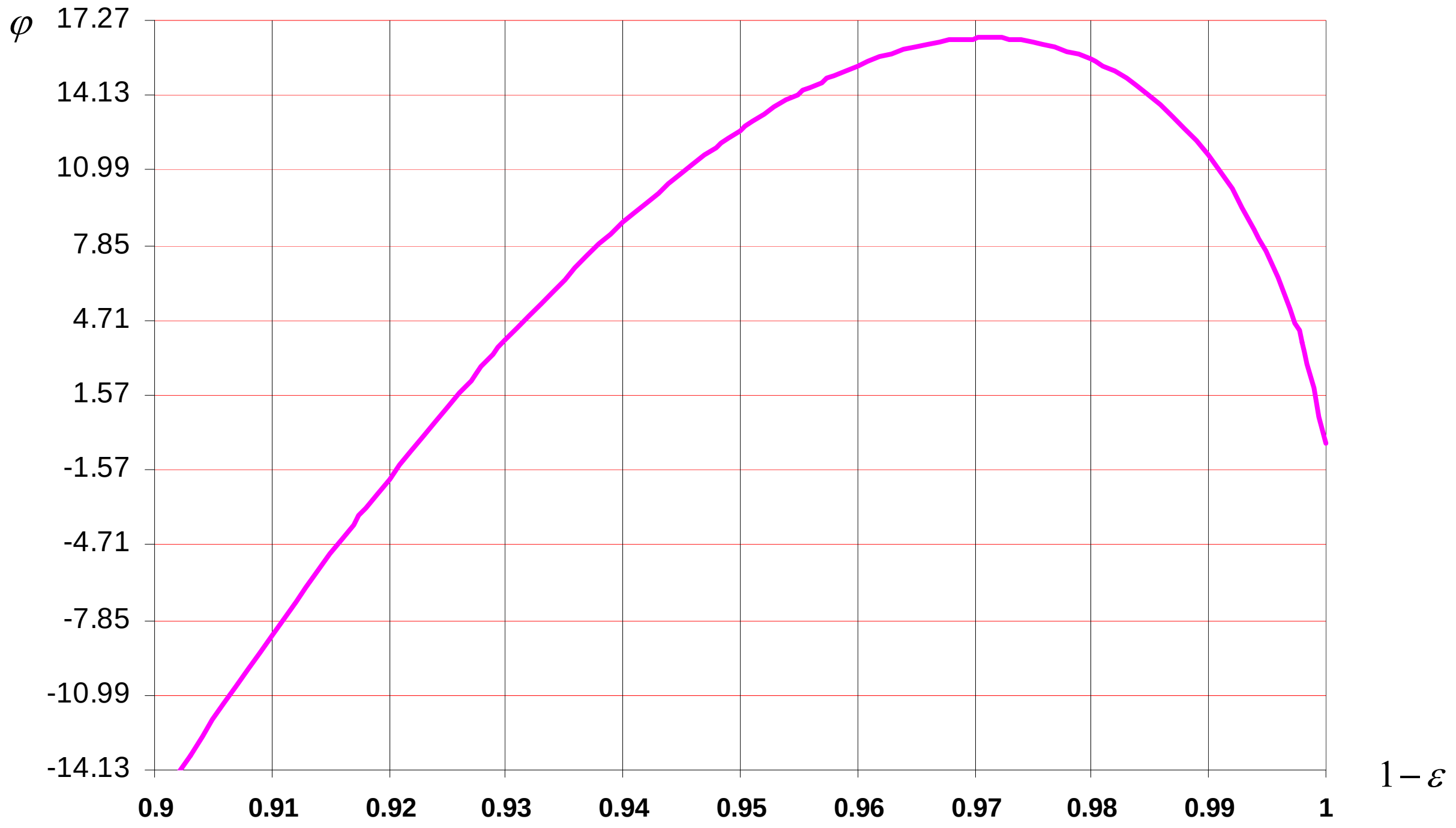

Figure 4. Phase behavior at $\alpha = 150$.

Now we proceed to specific results of calculations. For $\alpha \ll 1$, it is established in [14] - [18] that the energy spectrum must be close to hydrogen-like.

In addition, in [16], relativistic corrections to the hydrogen-like spectrum are calculated.

As a result, for $\alpha \ll 1$, the spectrum of a spin-half particle in the Schwarzschild field is given by the formula

$$\varepsilon_n = 1 - \frac{\alpha^2}{2n^2} - \frac{3\alpha^4}{n^4}\left\{ \frac{n}{l+\frac{1}{2}} \left[ \left(1 - \frac{1}{3}\delta_{l0}\right) + \frac{1}{12}\left(1-\delta_{l0}\right)\left(\frac{1}{\kappa} + \frac{2}{l(l+1)}\right)\right] - \frac{5}{8}\right\}. \quad (41)$$

Expression (41) shows that, as distinct from the hydrogen atom, the gravitational field removes degeneration of levels with the same $j$, but different $l$.

Table 1 shows analytical values of $1-\varepsilon_n^{an}$ for some values of $n$ and $\kappa$ obtained from expression (41) for $\alpha = 0,01; 0,05; 0,1$. The table 1 also shows similar values of $1-\varepsilon_n^{num}$ determined by numerical calculations. To within fractions of a percent, we see a close agreement between numerical and analytical values of $\varepsilon_n$.

Table 1. Numerical and analytical values of $1-\varepsilon_n$ for $\alpha = 0,01; 0,05; 0,1$.
(analytical values in ordinary type, numerical values in bold type)

| | $n=1$, $\kappa=-1$, $j=1/2$, $l=0$ | $n=2$, $\kappa=-1$, $j=1/2$, $l=0$ | $n=3$, $\kappa=-1$, $j=1/2$, $l=0$ | $n=2$, $\kappa=+1$, $j=1/2$, $l=1$ | $n=3$, $\kappa=+1$, $j=1/2$, $l=1$ |
|---|---|---|---|---|---|
| α = 0.01 | 5.0021E-05<br>**5.0040E-05** | 1.25038E-05<br>**1.2427E-05** | 5.5568E-06<br>**5.5133E-06** | 1.2502E-05<br>**1.2428E-05** | 5.5561E-06<br>**5.5139E-06** |
| α = 0.05 | 1.26633E-03<br>**1.2675E-03** | 3.14893E-04<br>**3.1487E-04** | 1.3967E-04<br>**1.3947E-04** | 3.1359E-04<br>**3.1316E-04** | 1.3929E-04<br>**1.3899E-04** |
| α = 0.1 | 5.2125E-03<br>**5.2866E-03** | 1.288E-03<br>**1.2970E-03** | 5.6805E-04<br>**5.7021E-04** | 1.2674E-03<br>**1.2691E-03** | 5.6189E-04<br>**5.6197E-04** |

| | $n=2$, $\kappa=-2$, $j=3/2$, $l=1$ | $n=3$, $\kappa=-2$, $j=3/2$, $l=1$ | $n=3$, $\kappa=+2$, $j=3/2$, $l=2$ | $n=3$, $\kappa=-3$, $j=5/2$, $l=2$ |
|---|---|---|---|---|
| α = 0.01 | 1.2501E-05<br>**1.2486E-05** | 5.5561E-06<br>**5.5134E-06** | 5.5558E-06<br>**5.5139E-06** | 5.5558E-06<br>**5.5729-06** |
| α = 0.05 | 3.134E-04<br>**3.1327E-04** | 1.3923E-04<br>**1.3883E-04** | 1.3904E-04<br>**1.3870E-04** | 1.3902E-04<br>**1.3904E-04** |
| α = 0.1 | 1.2643E-03<br>**1.2606E-03** | 5.6086E-04<br>**5.5911E-04** | 5.58E-04<br>**5.5746E-04** | 5.5769E-04<br>**5.5772E-04** |

As $\alpha$ increases, the energy spectrum begins to deviate from the hydrogen-like one.

Fig. 5 shows energy spectra for the three lowest energy levels in the state with zero orbital momentum ($1S_{1/2}$, $2S_{1/2}$, $3S_{1/2}$ ). As $\alpha$ increases, the binding energy $1-\varepsilon$ grows monotonically.

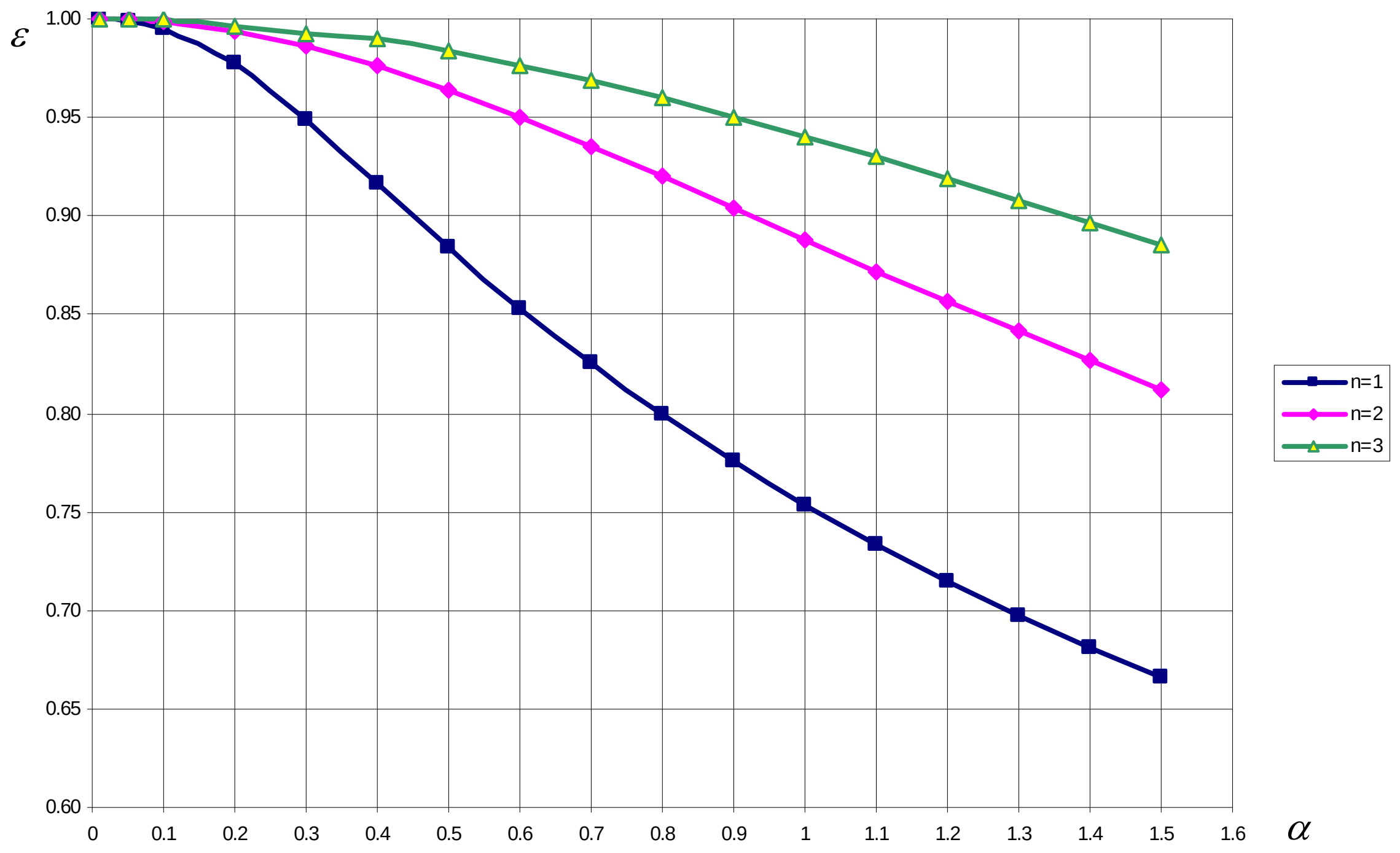

Figure 5. Energy spectra for the $1S_{1/2}$ $2S_{1/2}$ $3S_{1/2}$ states.

However, with a further increase in the coupling constant $\alpha$, the magnitude of the lowest energy level drops. For example, at $\alpha_{cr} \approx 2.65$, we observe a drop to $\Delta E \approx 0.35m$ shown in Fig. 6. With a further increase in $\alpha$, the lower energy level monotonically decreases. Note that all the levels above the $1S_{1/2}$ state also vary similarly as $\alpha$ increases. The presence of discontinuities in energy levels at certain values of the coupling constant is regular, and it is attributed to the behavior of $\varphi(\varepsilon)$ ($\varphi^{\max} = N\dfrac{\pi}{2}$ at $\varepsilon = \varepsilon^{cr}$).

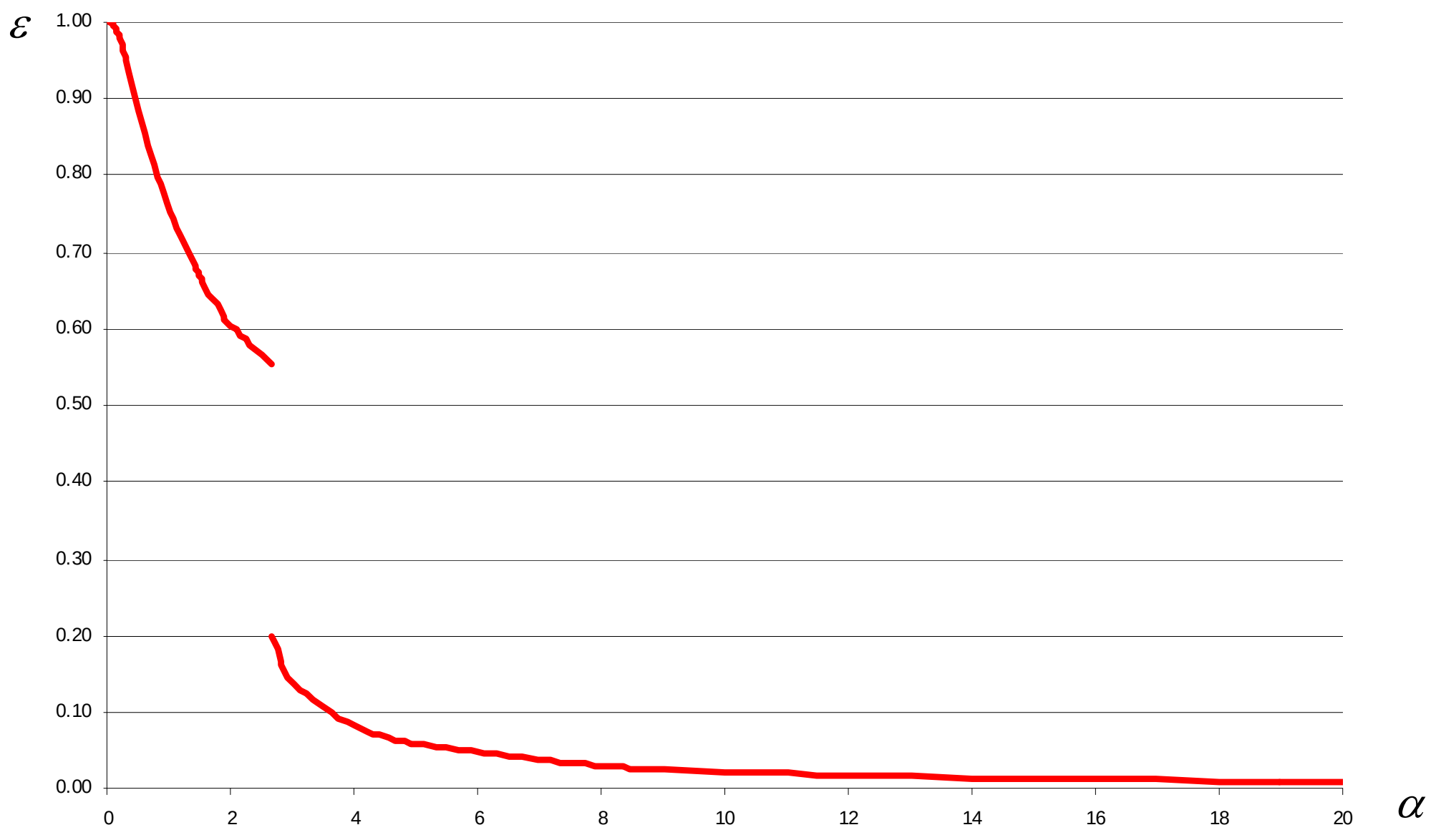

Figure 6. The function $\varepsilon(\alpha)$ for the $1S_{1/2}$ states.

The analysis of energy spectra obtained by numerical solution of system (29) and Eqs. (32), (35) for different values of $\kappa$ revealed a number of unusual features. Fig. 7 shows energy eigenvalues as a function of $\alpha$ for different values of $\kappa$. The plots at $\kappa = -2$ and $\kappa = 4$ display an unexpected behavior. These functions have a minimum at $\alpha \approx 0.35$ and $\alpha \approx 0.77$, respectively.

Fig. 8 shows the functions $\varphi(\varepsilon)$ for $\kappa = -2,\ \alpha = 0,4$ and $\kappa = -2,\ \alpha = 0,5$. We see that, as opposed to the phase behavior in most of the calculations done, in this case the phase maximum at $\alpha = 0,4$ is higher than in the calculation with $\alpha = 0,5$. As a result, the lowest energy level at $\alpha = 0,4$ is noticeably lower than at $\alpha = 0,5$. Further, at least up to $\alpha \approx 1$, the function $\varphi(\varepsilon)$ behaves in a standard way. A similar behavior of the function $\varphi(\varepsilon)$ at $\alpha = 0,8 \div 0,9$ is observed in the calculations with $\kappa = 4$.

Note that at least in the range of $\alpha \in (0.1 - 1)$ the lowest energy level is the state with the highest calculated total momentum (*j=9/2, l=4*), whereas the state (*j=7/2, l=4* ) lies much higher than the state (*j=7/2, l=3*), and at $\alpha = 1$ the energy difference between these two states reaches almost $0.8m$. Approximately the same splitting is observed between the levels with the same *l,* but different spin orientation. For example, at $\alpha = 1$, the splitting between the $4f_{5/2}$ and $4f_{7/2}$ level equals 0.76*m*.

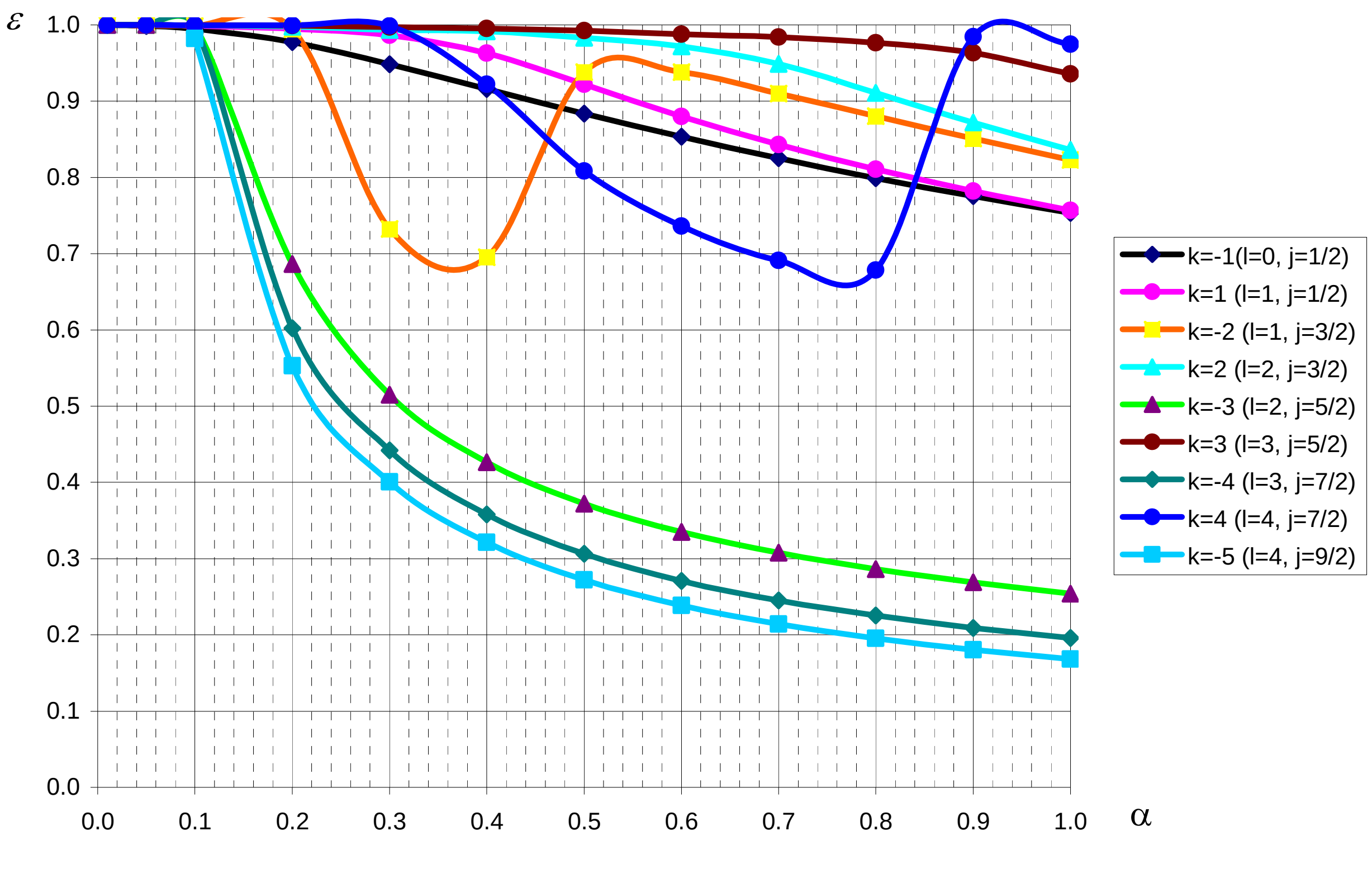

Figure 7. Functions $\varepsilon(\alpha)$ for the lowest energy levels and different values of $\kappa$.

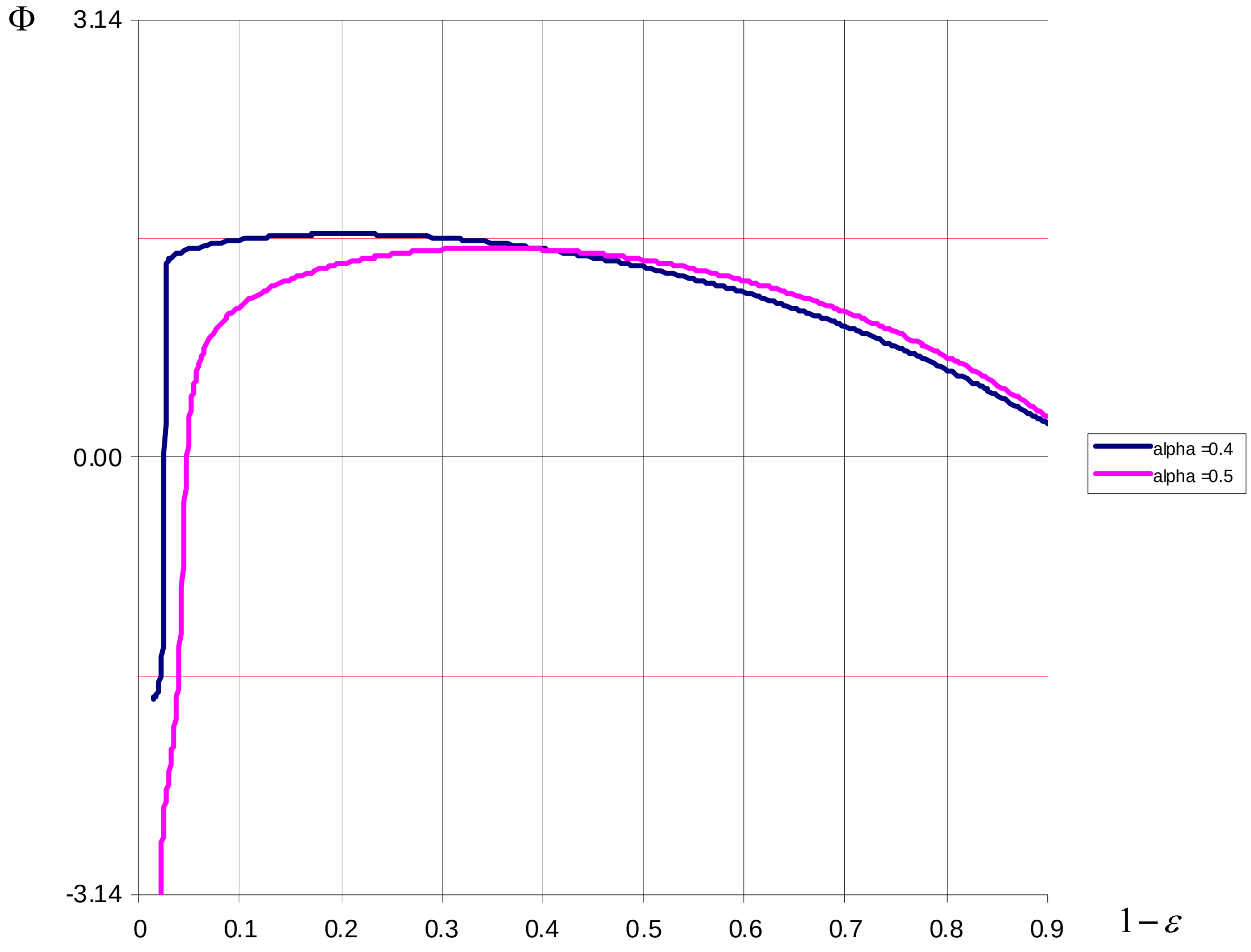

Figure 8. Behavior of phase curves at $\alpha = 0,4, \kappa = -2$ and $\alpha = 0,5, \kappa = -2$ .

Fig. 9 shows the growth in the number of tightly coupled $S_{1/2}$-states in the range of $\varepsilon = (0,1 \div 0)$ with increase in $\alpha$ .

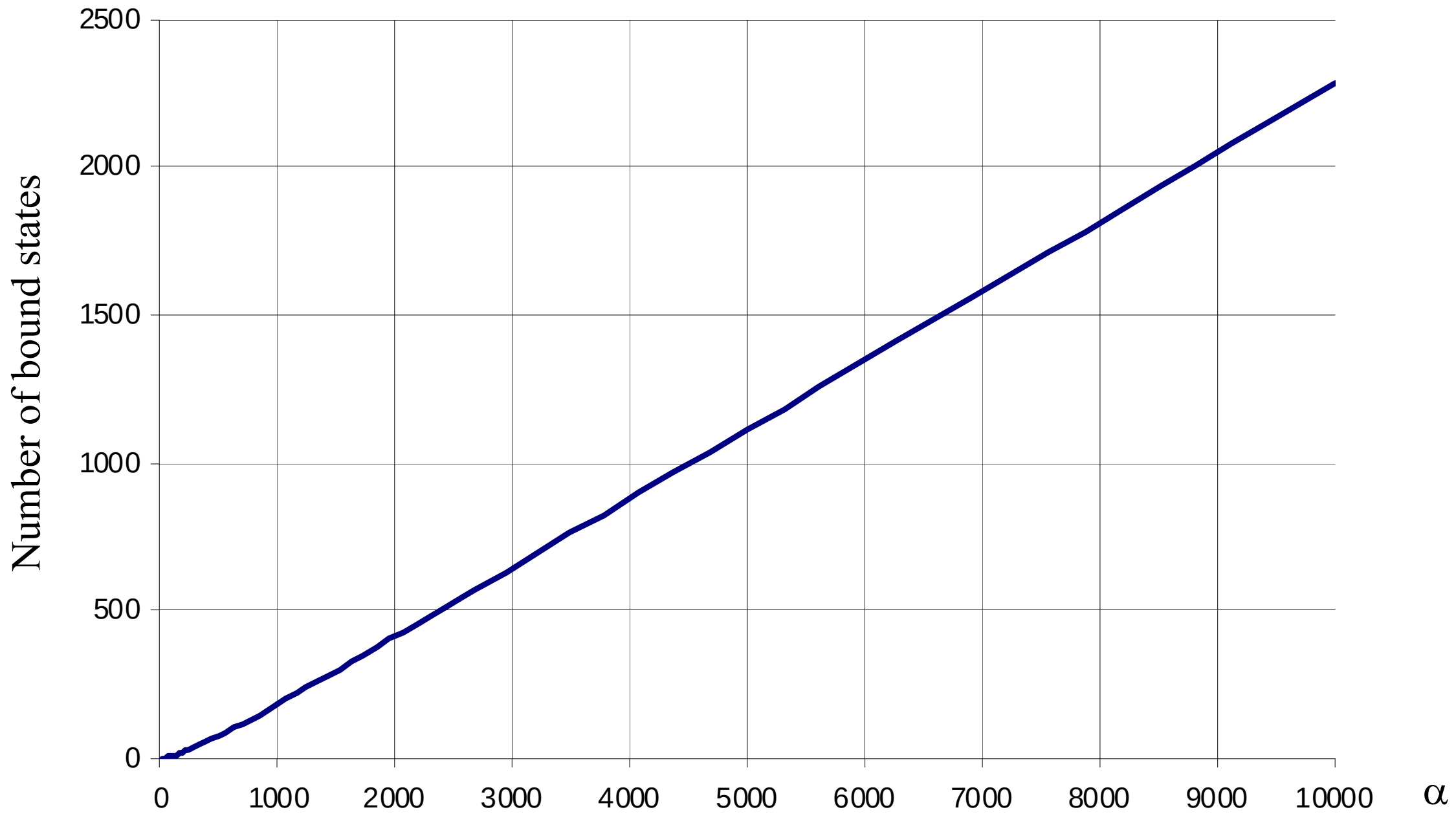

Figure 9. Number of $S_{1/2}$-states of the discrete spectrum in the range of $\varepsilon = (0,1 \div 0)$ as a function of $\alpha$ .

### 3.3.2.2 Radial wave functions

The time component of the Dirac current density in the $\eta$-representation for the Schwarzschild field is expressed as (see, (17))

$$j^0=\psi_\eta^+\psi_\eta=\frac{1}{\rho^2\left(1-\frac{2\alpha}{\rho}\right)}\left(f^2(\rho)+g^2(\rho)\right)\xi^+(\theta)\xi(\theta). \tag{42}$$

The $j^0$ component has simple physical interpretation: it is the probability density measured from a distant observer.

The quantity

$$J^0=\frac{1}{\left(1-\frac{2\alpha}{\rho}\right)}\left(f^2(\rho)+g^2(\rho)\right)=\frac{s+1}{s}\left(f^2(s)+g^2(s)\right) \tag{43}$$

for a distant observer is the probability density in a spherical shell of radius $\rho$.

The quantities $j^0$ and $J^0$ grow indefinitely as $\rho\to 2\alpha \quad (r\to r_0)$.

Similarly to an introduction of the boundary condition (38) at the determination of the energy spectrum of bound states (Subsection 3.3.2.1), when dealing with wave functions, we will choose $\rho_{\min}>2\alpha$ . In this paper for illustration we use $\rho_{\min}=2\alpha+1$.

Figs. 10, 11 show the functions $J^0(s)$ normalized to unity and calculated for some $S-$ and $P-$states.

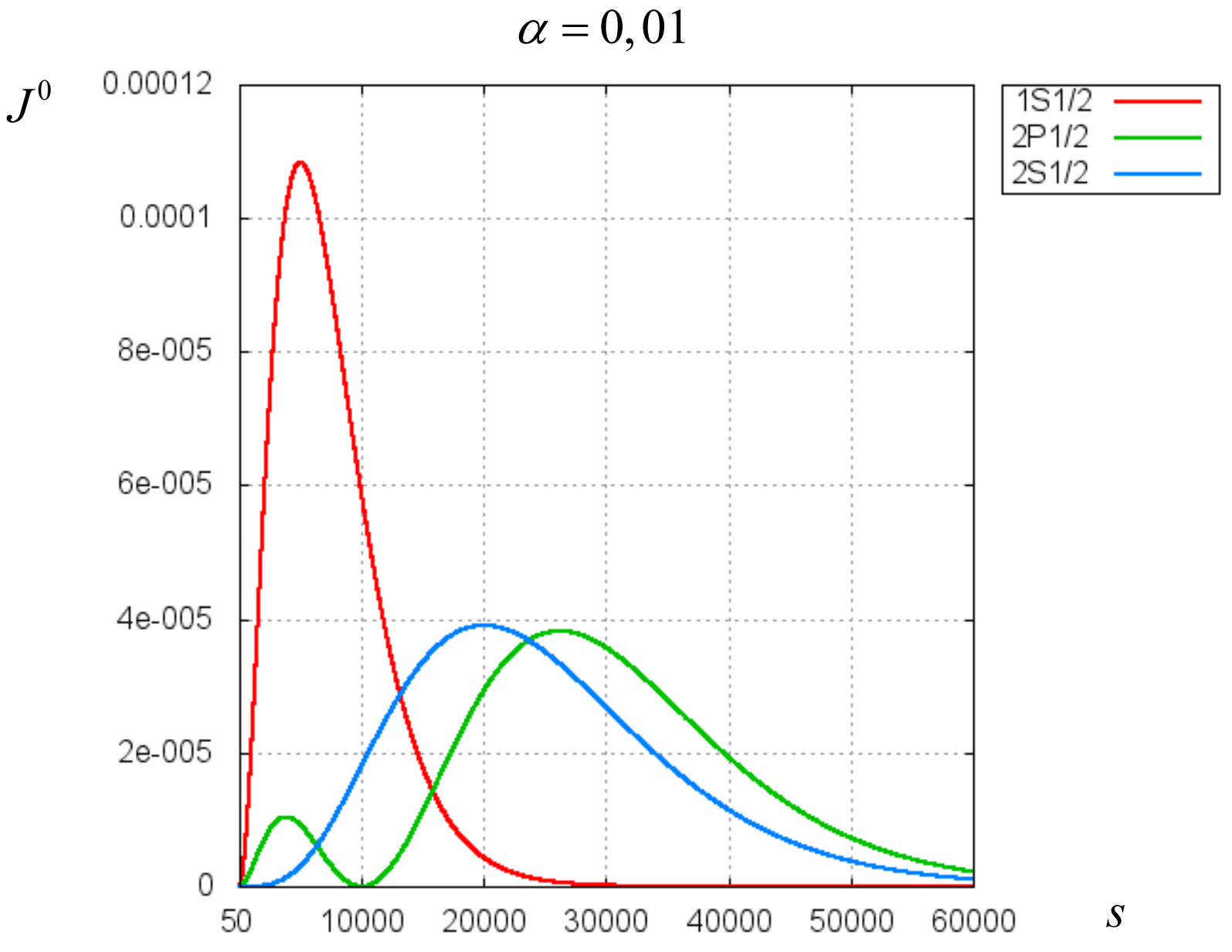

Figure 10. Normalized probability density as a function of $s=\frac{\rho}{2\alpha}-1$.

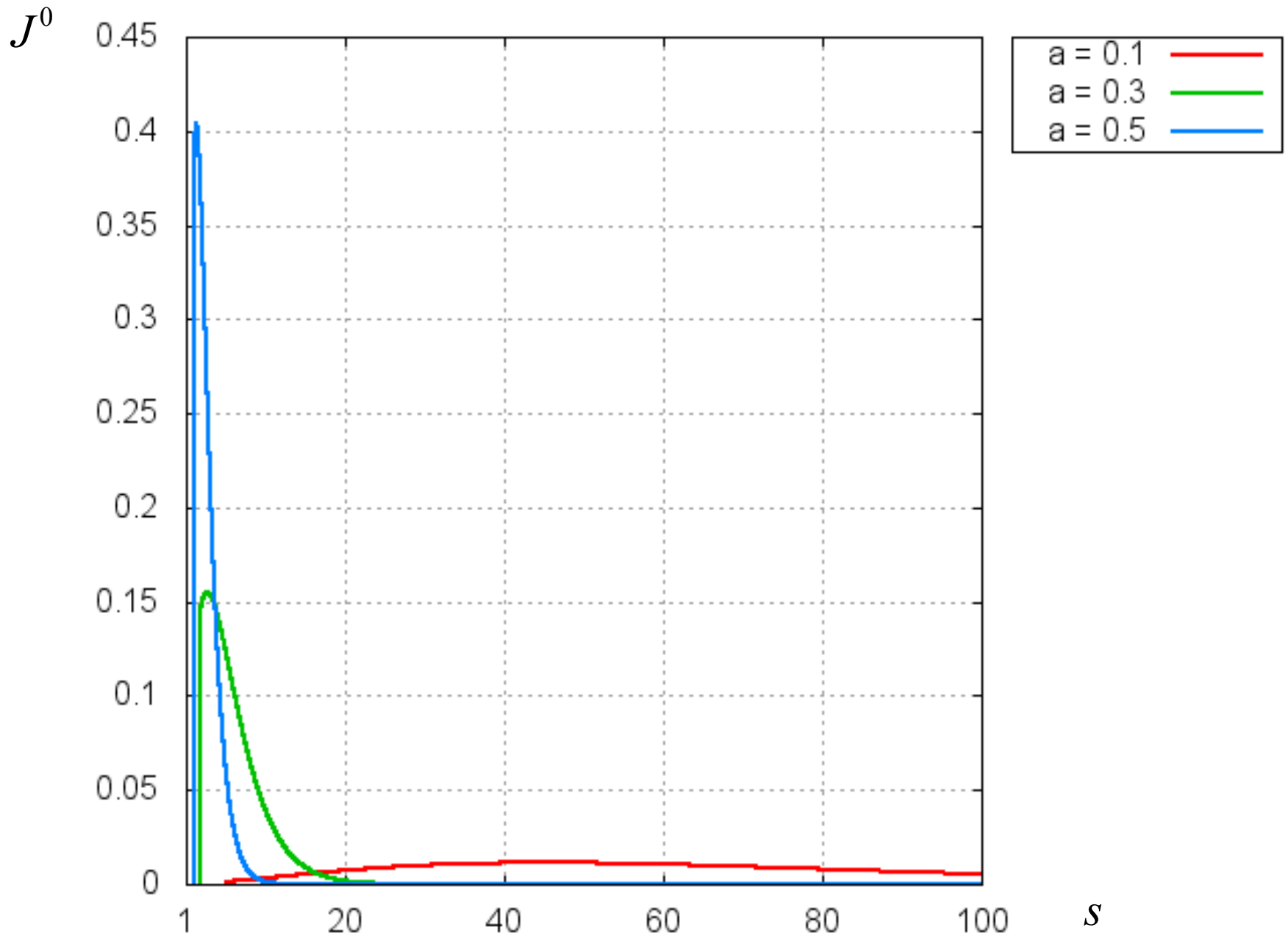

Figure 11. Normalized probability density for the $S_{1/2}$-states and different values of $\alpha$ as a function of $s = \frac{\rho}{2\alpha} - 1$.

Let us consider the problem of determination of mean radii with $\rho_{\min} = 2\alpha + 1$.

Apparently, the physical characteristic is excess of the mean radius over the gravitational radius: $\frac{\bar{\rho}}{2\alpha}$ or $\frac{\bar{r}}{r_0}$.

$$\bar{\rho} = \frac{\int\limits_{\rho_{\min}}^{\infty} \frac{\rho}{\left(1 - \frac{2\alpha}{\rho}\right)} \left(f^2(\rho) + g^2(\rho)\right) d\rho}{\int\limits_{\rho_{\min}}^{\infty} \frac{1}{\left(1 - \frac{2\alpha}{\rho}\right)} \left(f^2(\rho) + g^2(\rho)\right) d\rho} = 2\alpha \frac{\int\limits_{\frac{1}{2\alpha}}^{\infty} \frac{(s+1)^2}{s} \left(f^2(s) + g^2(s)\right) ds}{\int\limits_{\frac{1}{2\alpha}}^{\infty} \frac{s+1}{s} \left(f^2(s) + g^2(s)\right) ds}. \quad (44)$$

We see that for $\alpha \gg 1$ the corresponding mean radius for all states with various quantum numbers $n, j, l$ is close to the “event horizon” $(\bar{\rho} \to 2\alpha)$.

At $\alpha \ll 1$, the ratio $\frac{\bar{\rho}}{2\alpha}$ increases significantly and coincides with the ratio $\frac{\bar{r}}{l_c}$ in the hydrogen atom (see, e.g., [26]).

Table 2 shows numerical values of $\frac{\bar{\rho}}{2\alpha}$ for some $S-$ and $P-$states.

Table 2. Numerical values for $\frac{\bar{\rho}}{2\alpha}$ at $\alpha = 0,01$.

| | $1S_{1/2}$ | $2S_{1/2}$ | $2P_{1/2}$ |
|---|---|---|---|
| $\frac{\bar{\rho}}{2\alpha}$ | 150 | 600 | 500 |

Fig. 12 also represents the function $\frac{\bar{\rho}(\alpha)}{2\alpha}$ for the $1S_{1/2}$-state showing the values of $\bar{\rho}$ in the intermediate range of variation of $\alpha$.

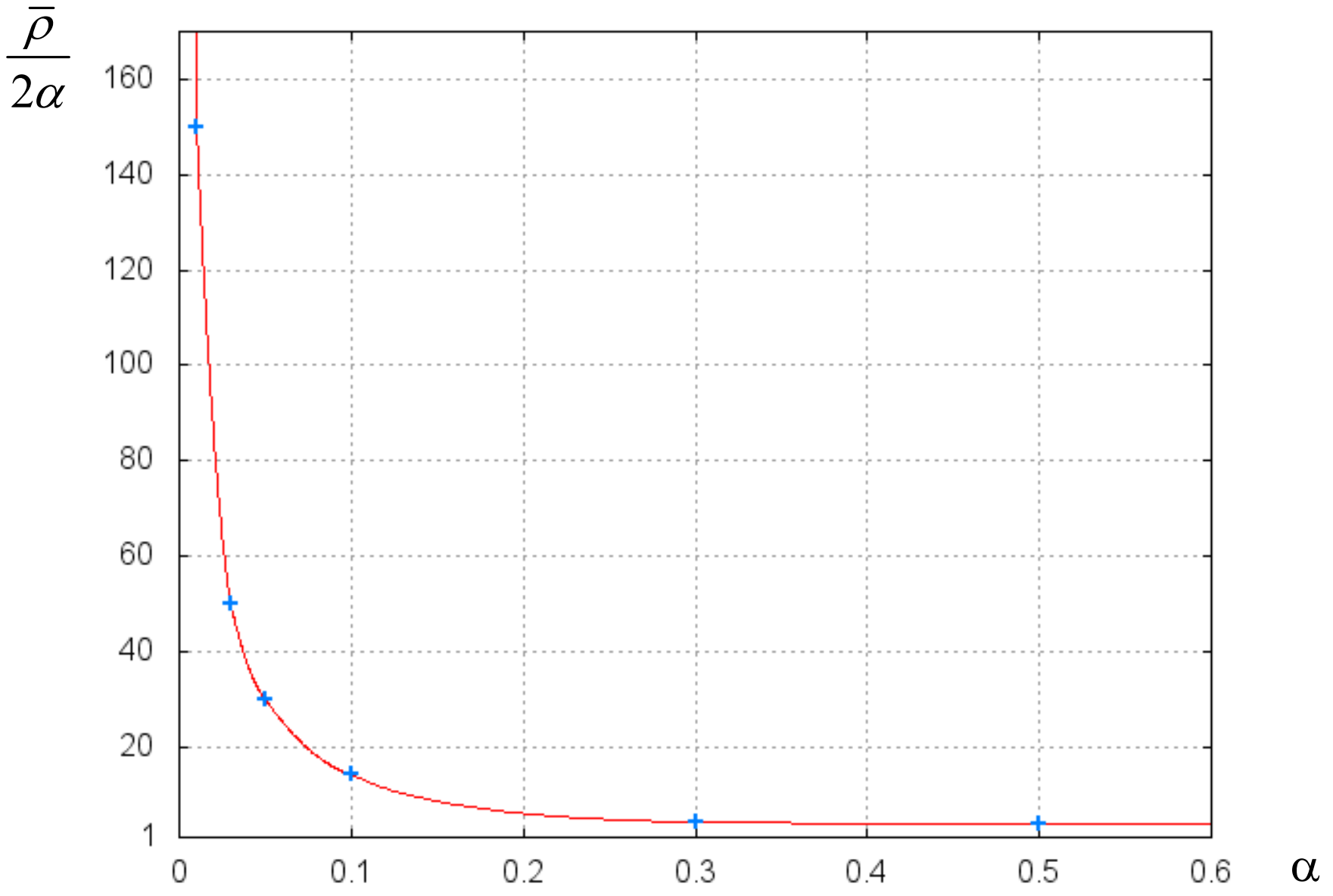

Figure 12. Deviation of $\bar{\rho}(\alpha)$ from the "event horizon" radius $2\alpha$ for the $1S_{1/2}$-state

## 4. Discussion

The paper shows the possibility of existence of stationary bound states of spin-half particles with a real energy spectrum in the Schwarzschild gravitational field of any strength.

The discrete spectrum is obtained in numerical calculations using self-conjugate Hamiltonian (5) with a flat scalar product of wave functions and boundary condition (26) providing zero value of the component of Dirac current density at $\rho = \rho_{\min}$, where $\rho_{\min} > 2\alpha$.

Note that in such a statement the Dirac quantum mechanical particle does not cross the "event horizon". The particle's wave function is determined on the interval $r \in [r_{\min}, \infty)$, where $r_{\min} > r_0$.

The numerical calculations allow us to draw the following conclusions:

1. For $\alpha \ll 1$, the energy spectrum is close to the hydrogen-like spectrum $\left(E_n \simeq m\left(1-\frac{\alpha^2}{2n^2}\right)\right)$.
2. Numerical relativistic corrections to the hydrogen-like spectrum for $\alpha = 0,01; 0,05; 0,1$ are close to the corrections determined analytically in [16].
3. As $\alpha$ increases, the energy of the bound particle decreases noticeably.
4. Energy levels undergo discontinuous variation when they reach their individual value of $\alpha_{cr}$.
5. Analysis of the spin-orbital interaction (calculations with different values of $\kappa$) confirms the conclusions made in [16], [24] that degeneration of energy levels with the same $j$ in the gravitational field is cancelled as distinct from energy levels in the Coulomb field.

   For $\alpha > 0,1$, calculations with $|\kappa| > 1$ reveal high splitting of energy levels with the same $j$ but different $l$, and also with the same $l$ but different projections of the particle spin $\left(j = l \pm 1/2\right)$.

   At $\alpha = 1$, for the energy levels with $j = 7/2, l = 3$ and $j = 7/2, l = 4$ this splitting becomes as high as $\Delta E \simeq 0,8m$. Nearly the same splitting at $\alpha = 1$ occurs for the levels with the same $l$: $4f_{5/2}$ and $4f_{7/2}$.
6. For $\alpha \gg 1$, the number of tightly coupled states in the range $\varepsilon \sim 0$ increases noticeably with increase in $\alpha$.
7. At $\alpha \ll 1$, the ratio of the mean radius of the bound Dirac particle to the gravitational radius is close to the ratio $\bar{\rho} = \frac{\bar{r}}{l_c}$ for the hydrogen atom; at $\alpha \gg 1$, $\bar{\rho} \to 2\alpha$ или $\bar{r} \to r_0$, that is, at high coupling, bound Dirac particles are "pushed" toward the gravitational radius (“event horizon”).

The existence of stationary bound states for Dirac particles is also possible for the Reissner-Nordström [2], Kerr [3] and Kerr-Newman [4] metrics, and the authors are going to discuss this in their subsequent papers.

The results of the work allow us to suppose that there exists a new type of collapsars. These collapsars are:

- inert (Dirac particles cannot pass through the “event horizon”);
- have no Hawking radiation property [25] (Hawking radiation requires that there is a wave function (Dirac field operators) under the “event horizon” [28] - [35]);

- provide for the existence of stationary bound states of spin - half particles for any values of the gravitational coupling constant $\alpha$ .

From the standpoint of cosmology, the new type of nonradiating relict collapsars can manifest themselves at small values of the gravitational coupling constant $\alpha \ll 1$ only through gravitation. Thus, they are good candidates as carriers of "dark matter". In the wide range of admissible masses, there can exist collapsars of a new type with the masses of hypothesized WIMP particles, which are treated as representatives of "dark matter" in numerous scenarios of the expansion of the universe.

On the contrary, the nonradiating relict collapsars with $\alpha \gg 1$, forming a "nested doll" of uncharged Dirac particles pushed towards the "event horizon" $\left(\bar{r} \to r_0\right)$, can play an important role in the formation of planets, stars etc.

Thus, the results of this study can lead to revisiting some concepts of the standard cosmological model related to the evolution of the universe and interaction of collapsars with surrounding matter.

## Acknowledgement

The authors would like to thank Prof. P.Fiziev for the useful discussions and A.L. Novoselova for the substantial technical help in the preparation of this paper.